\documentclass[10pt,notitlepage]{article}
\usepackage{graphicx}
\usepackage{natbib}
 \usepackage{amssymb}
\usepackage{amsbsy}
\usepackage{amsmath}
\usepackage{endnotes}
\usepackage{color}
\usepackage{caption}
\usepackage[multiple,flushmargin]{footmisc}
\usepackage[OT2,T1]{fontenc}
\usepackage{authblk}
 \usepackage[colorlinks=true]{hyperref}
 \hypersetup{urlcolor=rltgreen, citecolor=rltred, linkcolor=rltblue}
\usepackage{lineno}
\definecolor{rltred}{rgb}{0.5,0,0}
\definecolor{rltgreen}{rgb}{0,0.4,0}
\definecolor{rltblue}{rgb}{0,0,0.35}

\title{ \bf On the relationship between   cosmic rays, solar activity and powerful earthquakes. }

\author[1]{Kovalyov, M. (corresponding author)}
\author[$ $]{Kovalyov, S.}
\affil[1]{    email: mkovalyo@ualberta.ca}

\date{}
\begin{document}
\maketitle

\baselineskip20pt
\mathsurround=3pt
\parindent=25pt

\pagenumbering{gobble}
\pagenumbering{arabic} \setcounter{page}{1}

\noindent{\bf Abstract}. In this paper we analyze the correlation of cosmic rays intensity  to increases in seismic activity.  We also show that high-magnitude   earthquakes appear in group.  As a prequel, we discuss in \S1 naive visualization of the solar-cosmic ray interplay.

   \section*{  \centering{\S 1. Naive visualization of the solar-cosmic ray interplay.}}

 Earthquakes and volcanic eruptions have inspired fear since the very first days of man' presence on Earth.  To the same time date first attempts to predict them based on whatever our ancestors could see around: Sun, Moon, stars, weather, etc. and whatever they could not see: gods, deities, etc. Attempts to predict seismic activity still persists, mostly based on solar and lunar behavior. There are numerous publications on the existence of  correlation between the phases of the Moon and solar cycles and earthquakes and volcanic eruptions on Earth; many believe in it, their work is numerous and can be found by typing "Solar cycles earthquakes", "Moon phases earthquakes", "New Full Moon volcanic eruptions" or a similar expression into an Internet search engine. Others completely deny the existence of any such correlation, an good example is a recent paper \cite{lovej}, according to which solar-terrestrial triggering of earthquakes is insignificant. Which camp is right? Mostly likely each is somewhat right and somewhat wrong. Given the complexity of the events  contributing to seismic activity and our rather limited knowledge of them, it is rather naive to claim that one single event may cause seismic activity. Most likely, seismic activity is caused by a combination of factors, somewhat similarly to vehicular accidents usually caused by a combination of factors such as road conditions, weather, driver's condition, etc.  Sometimes one or two factors may dominate others, in which case the correlation is more visible;   however, when many factors are involved, the correlation between the cause and  result is barely seen.  Just quoted article \cite{lovej}  shows lack of outright
   statistical correlation between the solar activity and earthquakes, but an outright statistical correlation should not even be expected.   Drawing a conclusion from  lack of   a simple statistical correlation of the number and power of earthquakes with a single factor like the sunspot number  is like drawing a conclusion about  traffic accidents from  lack of a direct statistical correlation of traffic accidents with a single factor like drunk  drivers: not all  traffic accidents are caused by drunk  drivers, and many drunk drivers do not cause accidents. Yet we all know how dangerous   drunk drivers are, but we also know that accidents are caused not only by drunk drivers but also by drivers on medication, tired drivers, drunk pedestrians, slippery roads, thick fog, mechanical failures, improper maintenance, defective parts, etc.    What the article attests to is the complexity of seismic activity and its causes and how difficult it might be to claim a single factor as a cause of seismic activity.

\begin{figure}[!h]
  \centerline{\includegraphics[scale=.14]{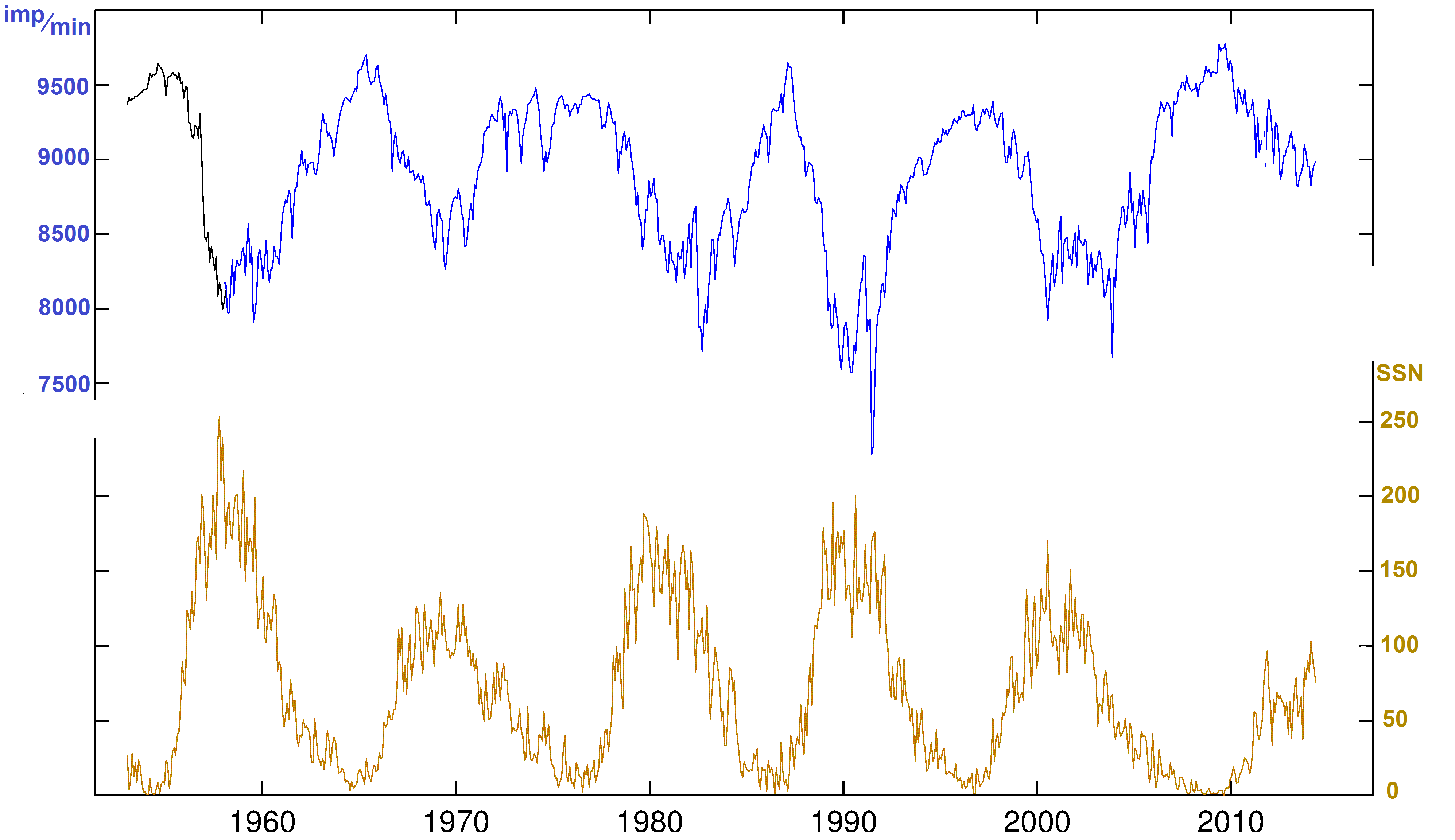}}
  \caption{ Cosmic ray intensity vs sunspot activity. The blue curve shows cosmic rays activity for the period  January 1, 1958 to May 1, 2014 according to  Moscow Neutron Monitor with the data taken from \protect \url{http://cr0.izmiran.ru/mosc/main.htm} . The black curve shows cosmic rays activity from January 1, 1953 to December 3, 1957 according to  Climax Neutron Monitor
Around of the station  with data taken from \protect \url{http://cr0.izmiran.ru/clmx/main.htm} and scaled to fit the blue curve, the scaling was done based on years 1958-2006 when both monitors were operational. The red curve shows monthly sunspot numbers with data taken from \protect \url{http://solarscience.msfc.nasa.gov/greenwch/spot_num.txt} .
 High sunspot activity correlates with low cosmic ray intensity  and vice versa.
  \hskip170mm  }
\label{fig:CRIvsSSN}
\end{figure}

   This article describes authors' observation that the seismic activity seems to show better correlation with cosmic rays, as measured by cosmic ray intensity and abbreviated by CRI, rather than with solar activity as measured by sunspot numbers and abbreviated by SSN.  The two are correlated as shown in \ref{fig:CRIvsSSN}; high SSN correlates with low CRI, Cosmic rays  are comprised of the extra-solar cosmic rays originating outside of the Solar System and the solar cosmic rays produced by the Sun.   Near Earth CRI is known to be
   modulated by cyclical solar activity with the average   cycle  of  about 10.85 -10.975 years long\footnote{The first number 10.85 is obtained by taking the time of 238.75 years between the first recorded maximum of solar activity in June 1761 and the last recorded maximum of solar activity in March, 2000 and dividing it by the number 22 of solar cycles between them;  the second number 10.975 is obtained by taking the time of  about 252 years and 5 months or 252.416667  years between the beginning of the first recorded solar cycle in August 1755 and the last recorded solar cycle in  January 2008 and dividing it by the number 23 of solar cycles between them.}. We shall refer to such cycles as {\it primary solar cycles}.  The physical mechanism of the solar activity is unknown but is currently believed to be solely of purely solar origin, with several theories attempting to explain it.   Cosmic rays actually are not rays at all but particles,  90\% of which  are  protons, 9\% are alpha particles.

\begin{figure}[!h]
  \centerline{\includegraphics[scale=.25]{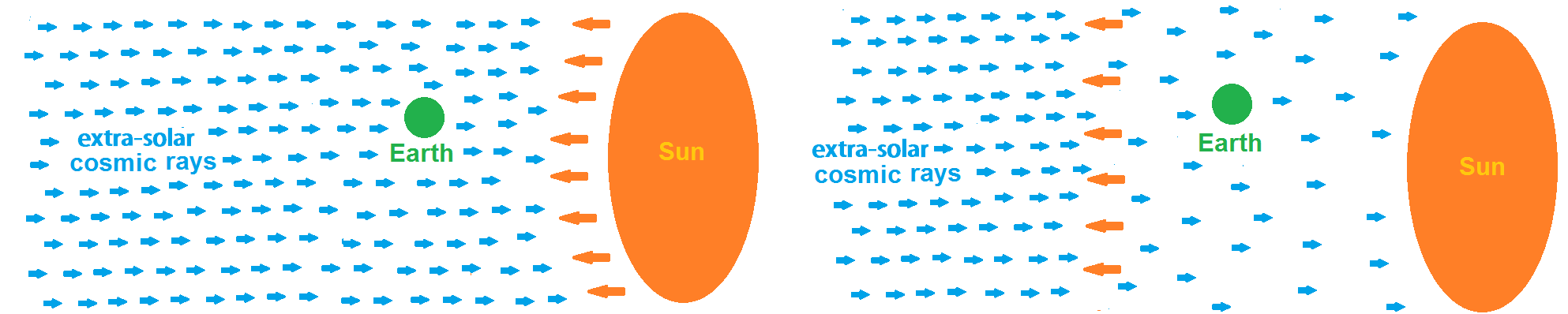} }
  \caption{ When SSN is at or just past  a minimum (as shown on the left), the solar wind is present but  weak and can deflect extra-solar cosmic rays only near the Sun; by the time it reaches the Earth it   loses its ability to deflect extra-solar cosmic rays. It takes several months   for the solar wind to gain sufficient strength    to deflect extra-solar cosmic rays for the time  it takes it to reach the Earth.  Hence the time lag between the maxima of CRI and the minima of SSN.  The solar wind produced when SSN is at or just past a maximum (as shown on the right) is so strong that it can deflect cosmic rays for much longer than the time required to reach the Earth.  As the solar wind passes the Earth and moves away from the Earth towards the interstellar space it continues to deflect extra-solar cosmic rays thus reducing the amount of cosmic rays reaching the Earth for several more months. Hence the time lag between the minima of CRI and maxima of SSN. This is just a simplified picture with cosmic rays shown as flowing in one direction, the cosmic rays are believed to come  not from one but from all directions.
  \hskip170mm  }
\label{fig:tm12}
\end{figure}

 A bit more careful analysis of Figure \ref{fig:CRIvsSSN}  shows that the maxima/minima of CRI  lag the minima/maxima of SSN by a few months.   The   time lag, explained in Figure  \ref{fig:tm12}, further further confirms the current paradigm that the solar activity modulates CRI.

\begin{figure}[!h]
  \centerline{\includegraphics[scale=0.27]{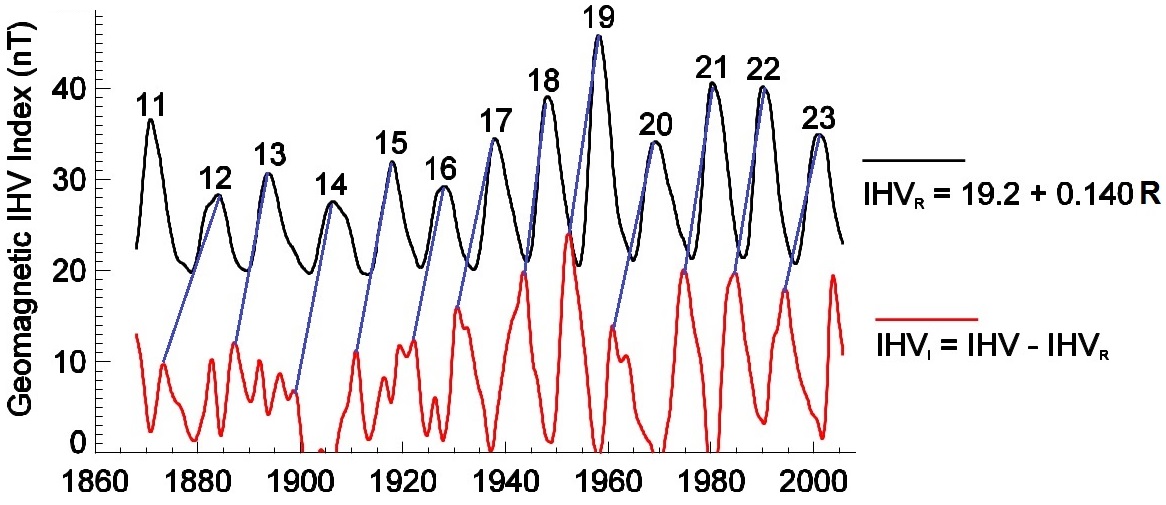} }
  \caption{ Solar activity, as measured by a long-term average of SSN,  vs geomagnetic activity, as measured by   $\rm IHV_I.$  Peaks in geomagnetic activity,  shown in red,  foretell solar maxima shown in  black  6-8 years in advance.  The numbers at the maxima of the solar cycles are the {\it solar cycle numbers.} The number  $\rm IHV_I$ is due to the  extra-solar cosmic rays, while the number  $\rm IHV_R$  is due to the cosmic rays from the Sun.  The solar activity is measured by $R = K (10g + s)$, where $g $is the number of sunspot groups and $s$ is the total number of distinct spots. Source:  \cite{hath}.
  \hskip170mm  }
\label{fig:hath}
\vskip3mm
  \centerline{\includegraphics[scale=0.40]{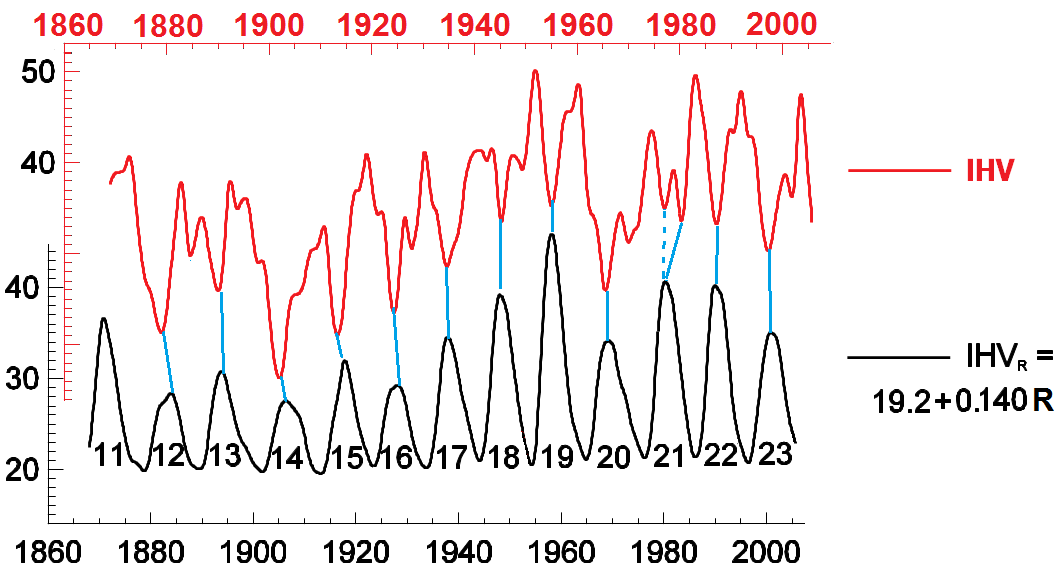} }
  \caption{ Solar activity, as measured by $\rm IHV_R,$ vs geomagnetic activity, as measured by $\rm IHV.$ The black graph with black axes and labels shows  $\rm IHV_R,$   the numbers at the base  of a solar cycles is the {\it solar cycle numbers}. The red graph with red axes and labels shows   IHV. Notice the red graph is shifted right by about 30 months.  The minima in geomagnetic activity, as measured by  IHV,  foretell solar maxima   about 20-30 months  in advance for all but one solar maximum, the minima of IHV are connected to the corresponding maxima of $\rm IHV_R$  by solid blue lines. The only exception is solar cycle 21, its maximum was only about one month after  the corresponding  minimum of IHV, yet there was one more local minimum of IHV preceding the solar maximum and it was  20-30 months ahead of the solar maximum.  Here $R = K (10g + s)$, where $g $is the number of sunspot groups and $s$ is the total number of distinct spots.  Source:  \cite{hath2}.
  \hskip170mm  }
\label{fig:hath2}
\end{figure}

Recent work  \cite{hath}, however, suggests that not only SSN modulates  CRI but also CRI affects solar activity and SSN. The authors of   \cite{hath} looked at the records of geomagnetic activity stretching back almost 150 years and noticed that the  geomagnetic activity precedes the solar cycle by about  6-8 years.  Although their original  prediction   that the peak of the current solar  cycle would be     one of the most intense  since record-keeping began almost 400 years ago  failed spectacularly, it was further corrected to be in line with the rest of their predictions.  Their main idea, that the values    of the   maxima of a long-term average  R  of SSN  are correlated to the values of the preceding maxima of $\rm IHV_I$ is illustrated in  Figure \ref{fig:hath}.    Another correlation between the solar and geomagnetic activities is shown in Figure \ref{fig:hath2}, where the minima of IHV seem to follow the same pattern as the maxima of SSN.  A number of similar correlations is described in \cite{predic3, predic1, predic2},  of all such correlations according to \cite{predic3},   three may serve as the most reliable predictors; all three are based on geomagnetic activity near  solar   minima when the solar cosmic rays are at their maximum.

 How could geomagnetic activity predict solar activity? One may entertain  several possibilities: 1) the Earth affects the Sun and determines solar activity; 2) there is an invisible and undetectable, so far at least, component of solar activity that precedes visible and  detectable components of solar activity, its effects on  the Earth exhibit themselves prior to the visible effects of solar activity; 3) there is a third agent that affects both the Sun and Earth, its affects on Earth  show up earlier than its effects on the Sun due to Earth's much smaller size.  The first hypothesis is hardly believable mainly because the Earth is so much smaller than the Sun; the second hypothesis is also hard to believe. The third hypothesis is most likely with the third agent being cosmic rays of extra-solar origin.  That the cosmic rays are correlated with volcanic eruptions of a certain type of volcanoes was pointed out  in a recent paper \cite{cosmrayvolc}, although we  may disagree with the explanation provided in the article.  We will not try to conjecture any theories here as to how extra-solar cosmic rays may affect the Sun and Earth but merely show the correlation of seismic activity on Earth with  CRI.  The relationship between cosmic rays and solar/terrestiral activity is two-way, while cosmic rays affect Sun/Earth, the Sun's/Earth's magnetic field in turn deflects the flow of cosmic rays thus affecting their intensity in the solar system/near Earth.

      \begin{figure}[!h]
  \hskip2mm{\includegraphics[scale=0.64]{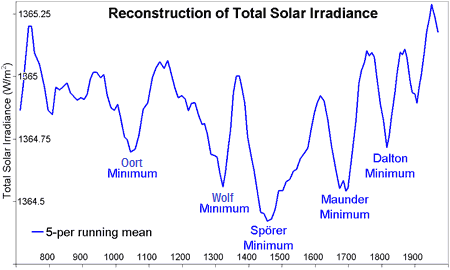} }
\label{fig:dela}
\vskip-21mm
  \hskip79mm{\includegraphics[scale=0.179]{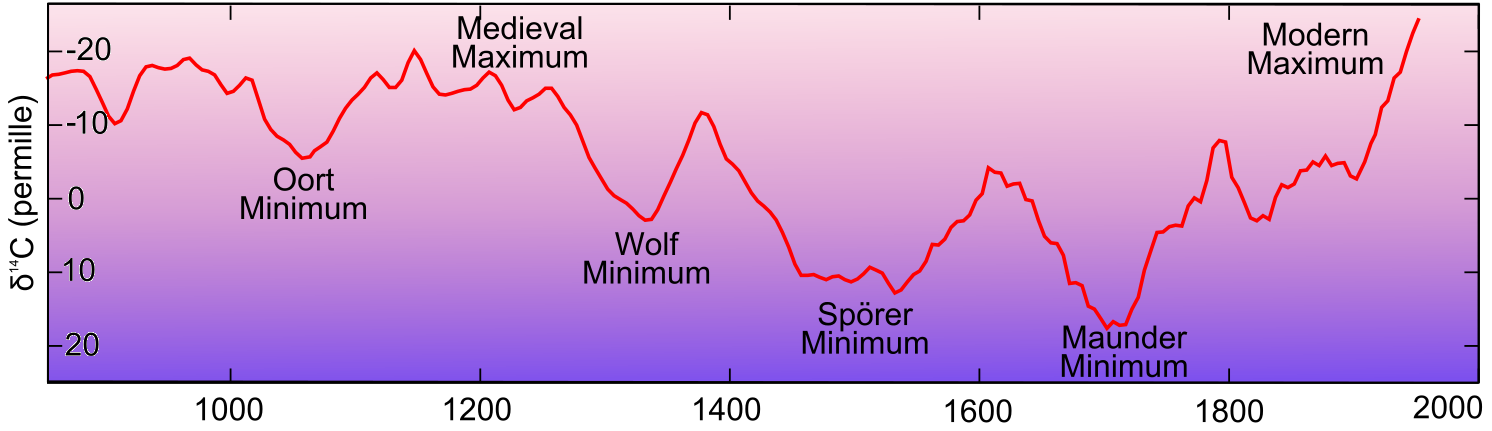} }
  \vskip-3mm\caption{The left graph shows solar activity for the past 1300 years as represented by total solar irradiance, reconstructed based on $^{10}B, $ the graph on the right shows solar activity for the past 1200 years,  reconstructed based on $^{14}C.$  Source:  \cite{delaygue},  \cite{c14}. }
\label{fig:carb}
\end{figure}
\begin{figure}[!h]
  \hskip2mm{\includegraphics[scale=0.65]{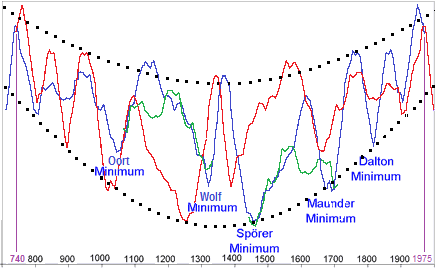} }
\label{fig:dela1}
\vskip-21mm
\hskip78mm{\includegraphics[scale=0.18]{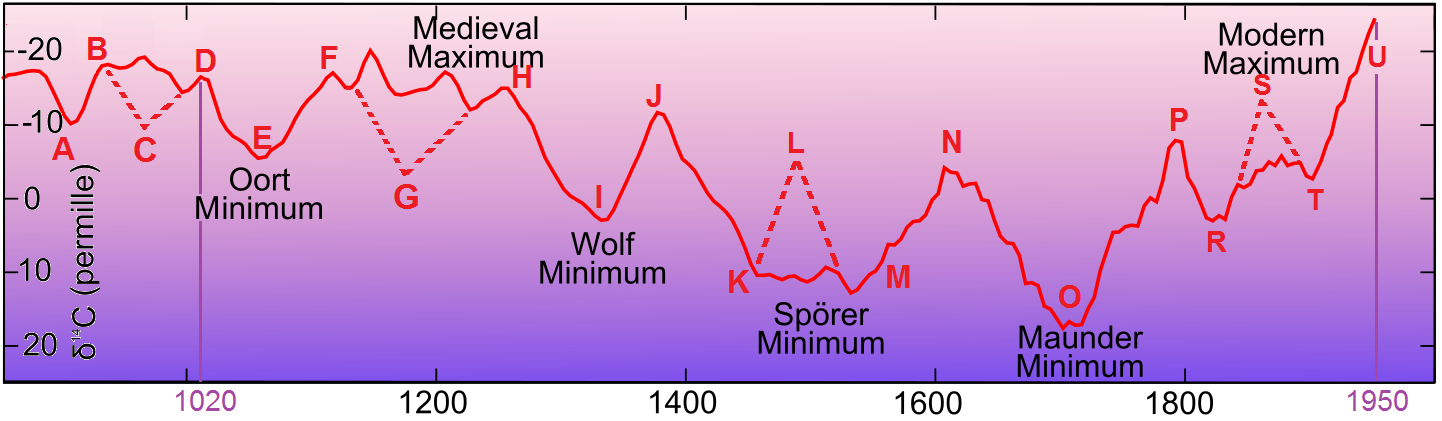} }
  \caption{ The blue curve in left graph is the same as in the left graph in Figure \ref{fig:carb}, the red curve is the horizontally flipped image of the blue curve; the green curve on the left is  part of the blue curve, the green curve on the right is  the horizontally flipped image of the green curve on the left,  the green curves are inserted to show that the period between the Dalton and Maunder Minima  solar cycles is actually two secondary solar cycles blended together, as is the period between the Wolf and Oort Minima. Thus the left graph shows periodicity of  about $\frac{1975-740}{11}\approx 112.2727$ years
The   graph on the right is the same as the    graph  on the right in Figure  \ref{fig:carb} with broken lines indicating activity that seemingly should have occurred but  did not; the graphs shows periodicity of  about $\frac{1950-1020}{8}\approx 116.25$ years. The absence of the solar activity's effect on Earth  shown by the dotted curves in the right  graph    is currently attributed to the changing strength of the Earth's own magnetic field; however, the change in the strength of the Earth's magnetic field itself might be due to the change in CRI, after all, cosmic rays is a current of electrically charged particles which generates a magnetic field that interacts with the currents of electrically charged liquid metal inside the Earth's liquid core responsible for the Earth's magnetic field.
  \hskip70mm  }
\label{fig:carb1}
\end{figure}

 The primary solar cycles of about  10.85 -10.975 years  are modulated by much longer {\it secondary  cycles}\footnote{These cycles are not to be confused with spectral cycles of average length of 87 years (Gleissberg cycles),  104 years, 150 years,  210 years (Suess-de Vries cycles), 506 years, etc.   }, two reconstructed version of which are shown in  Figure \ref{fig:carb}.    We use
  Figure  \ref{fig:carb1} to determine the average length of the secondary cycles, the graph on the left gives us 112.2727 years as the average value of the secondary period, while the graph on the right gives us 116.25 years as the average value of the secondary period. In the graph on the left we
    superimposed the  left  graph from Figure \ref{fig:carb}  shown in blue, with  its horizontally flipped image  shown in red, and two   green curves, the one on the left is  part of the blue curve, the one on the right is  the horizontally flipped image of the green curve on the left.  The green curves are inserted to illustrate that the period between the Sporer and Maunder Minima  is actually two secondary solar cycles blended together, as is the period between the Wolf and O\"{o}rt Minima.
According to  \cite{delaygue}, the Dome Fuji record features a secondary peak  between   1570 and 1600 AD, while \cite{mcc} also predicts the existence of a secondary maximum centered around year 1220 AD, thus confirming our hypothesis
that the periods between the Oort and Wolf Minima  and  between the Sp\"{o}rer and Maunder Minima contain   two rather than one secondary solar cycles merged together.     Since both the arithmetic  and geometric means of 112.2727 and 116.25    are correspondingly 114.2527103 and 114.25271, both the arithmetic and geometric means of the which are approximately 114.25271,  we will take it as  average length of a secondary cycle.
 The 114.25271-year length of secondary solar cycles is supported by \cite{scaf}  which discusses  an evidence of the existence of 60-year (about half of 114.25271 years) long cycles in nature.  The 114.25271-year length of the secondary solar cycle might also be the reason why  the Chinese Zodiac is comprised of 12 (about one tenth of 114.25271) years/signs and  why the Chinese time counting involves  60-year cycles.
  There are lots of periodic variations  affecting CRI, e.g periodic   variations  of the Earth's axial tilt,    of the Moon's declination,   of the gravitational pull by other planets, etc.; they certainly should  have been  taken into account but had not been, without them the 114.25271-year length of the period should be viewed as a 'naive' zero approximation.

There is currently no theory as to how solar cycles work, just some hypotheses.  To help us visualize  how secondary cycles might be generated in the solar-cosmic ray interplay, let us think of the Sun's surface as a  fluid at the boiling point,  part of it is in gaseous form and the rest is in liquid form.
The surface   is bombarded with fast moving (particles of) cosmic rays,  which, upon hitting the surface, do two things 1)  increase the temperature of the gaseous part of the fluid  above the boiling point allowing more and more fluid's particles to escape the Sun  as   solar wind; 2) change  part of the   fluid from liquid   into gaseous state.  Some parts of the gaseous state revert back to liquid while emitting energy in the form of solar irradiance. The process is similar to boiling water, with   sunspots and solar wind  being correspondingly the analogues of   bubbles   and steam.   The particles of   solar wind, moving much slower than cosmic rays,  deflect the   latter decreasing their number; with fewer and fewer  cosmic rays  hitting the surface, the temperature    drops, the sunspots become fewer and fewer and the solar irradiance and wind gradually decrease. With less solar wind, more and more cosmic rays hit the surface   raising its temperature again and the cycle repeats itself.    If, for whatever reason, the  intensity of cosmic rays drastically increases, the gaseous portion of the fluid  may   {\it steam out} leaving it smaller than its average size;  a more sizable portion of the energy of the cosmic rays then goes to the heat of vaporization   required to turn some of the liquid   into the gaseous state; the solar irradiance and   solar wind decrease as do the sunspots numbers. We shall call this phase of solar activity a {\it steam-out period}.  Once more liquid is changed into gaseous state, the solar output in the form of solar wind and solar irradiance increases, as does the number of sunspots.  The steam-out may occur over several cycles with the gaseous portion of the fluid  getting smaller and smaller with each consecutive cycle until it reaches its climax, from where the size of the gaseous portion begins to increase again.
 Figures \ref{fig:carb}-\ref{fig:carb1}  indicate  the steam-out period   between the Wolf and Maunder Minima, climaxing at the time of  Sp\"{o}rer Minimum, it followed    the Medieval Maximum when large amount of the gaseous portion of the fluid   must have evaporated from the surface of the Sun.   We would like to repeat that the current paragraph is not a scientific theory or even a hypothesis, it is merely a way to help us visualize the solar-cosmic ray interplay; a proper model must be taken into account not only thermodynamic but also electromagnetic aspects of  solar plasma as well as the motion of plasma within the Sun. Having said that, some kind of phase transition in solar plasma in response to changes in cosmic rays flow is most likely a large contributor to secondary, and possibly even primary, solar cycles.

 The interaction of cosmic rays with the Earth  is not just two-way but rather three-way: 1) cosmic rays affect the Earth   directly in a variety of ways; 2) cosmic rays affect the Earth indirectly by affecting solar activity; 3) the changes in the Earth's magnetic field, some due to    the flow of plasma  in the liquid core also responsible for seismic activity on Earth, affect the cosmic ray flow near Earth.  Thus the correlations between CRI and seismic activity described in the next section may have occurred through either one of these channels or a combination of either two or even all three of them.

 Where does the period of 114.25271 years come form? It is not clear; most likely, it is the average period of motion of the source(s) of cosmic rays, or at least, an important component of cosmic rays.
The thought that a significant part of cosmic rays comes from a single source  may appear  contrary to the observation of  cosmic rays coming from all directions.  Yet it is not. Just like on a  cloudy and foggy day the sunlight coming from the Sun  appears to arrive from all directions due to scattering by  clouds and fog;    cosmic rays coming  from a single source may appear to be coming  from all directions due to scattering by   interstellar and galactic gases,  electromagnetic and gravitational fields,  and, in no small part,     the solar wind.  The presence of a small anisotropy in cosmic rays flow is discussed in \cite{abb1, abb}.
If we are correct, then the events on Earth are strongly influenced by events very far, perhaps thousands of light years, away.

 \section*{  \centering{\S 2. Cosmic rays and seismic activity on Earth.}}

Let us start by looking at  earthquakes of magnitude $\geqslant 7.8$ since 1900, these are the most powerful earthquakes,  cr\`{e}me de la cr\`{e}me of natural disasters.  The cut-off point $7.8$ for the magnitude was selected to have sufficiently many earthquakes to draw conclusions yet not too many to be overwhelmed.
\begin{figure}[!h]
  \centerline{\includegraphics[scale=0.15]{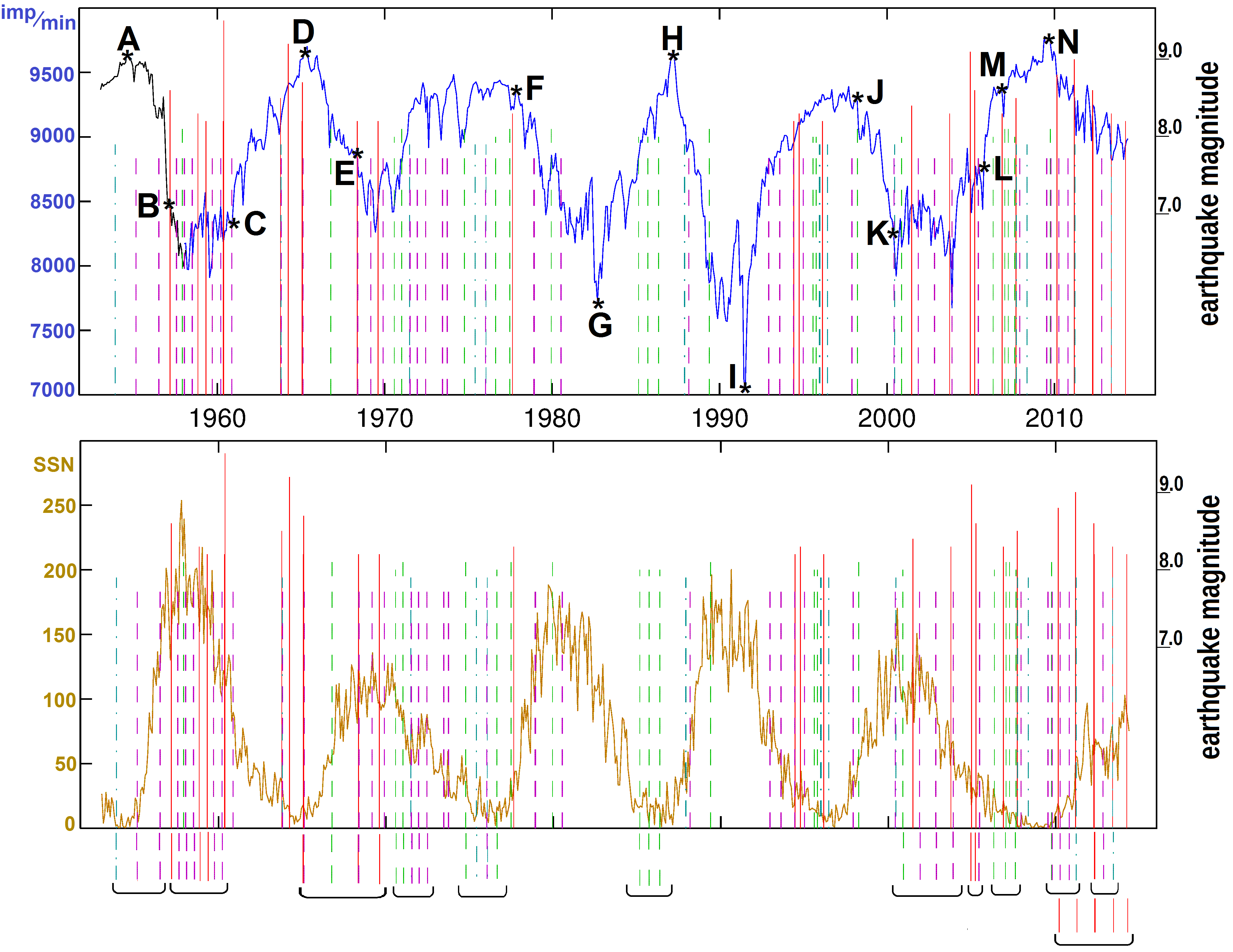} }
  \caption{ CRI and SSN vs earthquakes of magnitude $\geqslant 7.8$. Red solid vertical lines represent  earthquakes of magnitude $>8.1$,  green broken vertical lines represent  earthquakes of magnitude   $8.0$ and $8.1,$  green dot-dash vertical lines represent  earthquakes of magnitude   $7.9,$ and purple broken vertical lines represent  earthquakes of magnitude   $7.8.$ The blue and brown curves show  CRI and SSN as in Figure \ref{fig:CRIvsSSN}. Several  groups of earthquakes are shown under the main graph.  The list of the earthquakes was obtained by taking all earthquakes of magnitude $\geqslant 8.0$  from \cite{largestearthq8} and supplementing with the  magnitude $7.8 -7.9$ from \cite{earthusgs}. Although both web sites are  produced by USGS,
some of the earthquakes are assigned different magnitude and/or different date/time, the differences are usually insignificant although in some cases they are fairly considerable, e. g.  \cite{largestearthq8} as of January 20, 2015 states that the July 31, 1970 earthquake in Columbia was of magnitude 8.0 and the January 10, 1971 earthquake in Indonesia was of magnitude 8.1 while \cite{earthusgs} states that the magnitudes of the two earthquakes were correspondingly 7.5 and 7.7.   Sifting through the two lists, double entries,  relatively insignificant  aftershocks and foreshocks were removed, dates and magnitudes were adjusted based on additional sources.     }
\label{fig:CRIvsEarthq}
\end{figure}

In Figure \ref{fig:CRIvsEarthq} earthquakes of magnitude $\geqslant 7.8$ from 1958 to 2010  are superimposed on the graphs of CRI and SSN from Figure \ref{fig:CRIvsSSN}. One may notice that:

\noindent 1. The powerful earthquakes are much less frequent when the CRI  decreases sharply over a prolonged period of time, i.e  between A and B, D and E, F and G, H and I. There is only one earthquake between D and E; there are only three earthquakes between F and G;   there are only three earthquakes between H and I. Even the short-lived decrease between points J and K saw only one powerful earthquake.

\noindent 3. Of the six CRI cycles,  only two,  between points C and E and between  points H  and I had triangular shape, both cycles had a significantly smaller number of powerful earthquakes.  The average rate of change between points C and D, D and E,  and   H  and I  was larger than average.

\noindent 4. The density of powerful earthquakes is the highest during  time intervals when the average rate of increase is relatively small,  that is between points B and C, between points E and F, between points  K and L and past point N. Notice between points E and F the minimum almost instantly turned to a maximum and past point N the maximum almost instantly turned to a minimum.

\noindent 4. High magnitude earthquake often appear in almost periodic groups,  the higher is the density of earthquakes the more groups appear.

There are no reliable CRI records for the period 1900-1953, however there are reliable records of SSN which may be used as a proxy for CRI while keeping in mind that CRI  may lag SSN by  up to 18 months.

\begin{figure}[!b]
  \centerline{\includegraphics[scale=0.14]{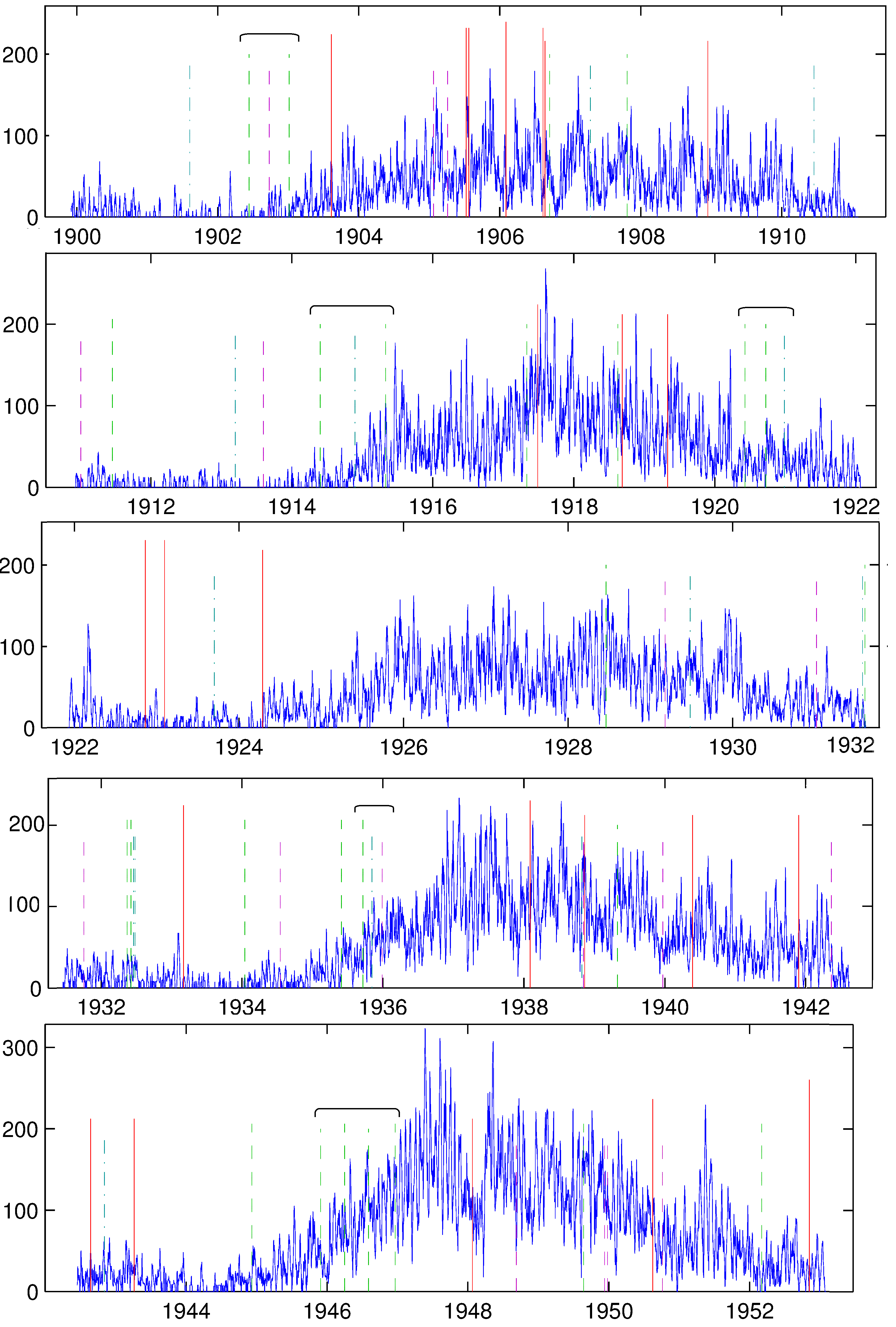} }
  \caption{ SSN vs earthquakes from January 1, 1900 to August 15, 1942. Notations are explained in the caption to Figure \ref{fig:SSNvsEarthq3}.
  \hskip170mm  }
\label{fig:SSNvsEarthq1}
\end{figure}

\begin{figure}[!t]
  \centerline{\includegraphics[scale=0.14]{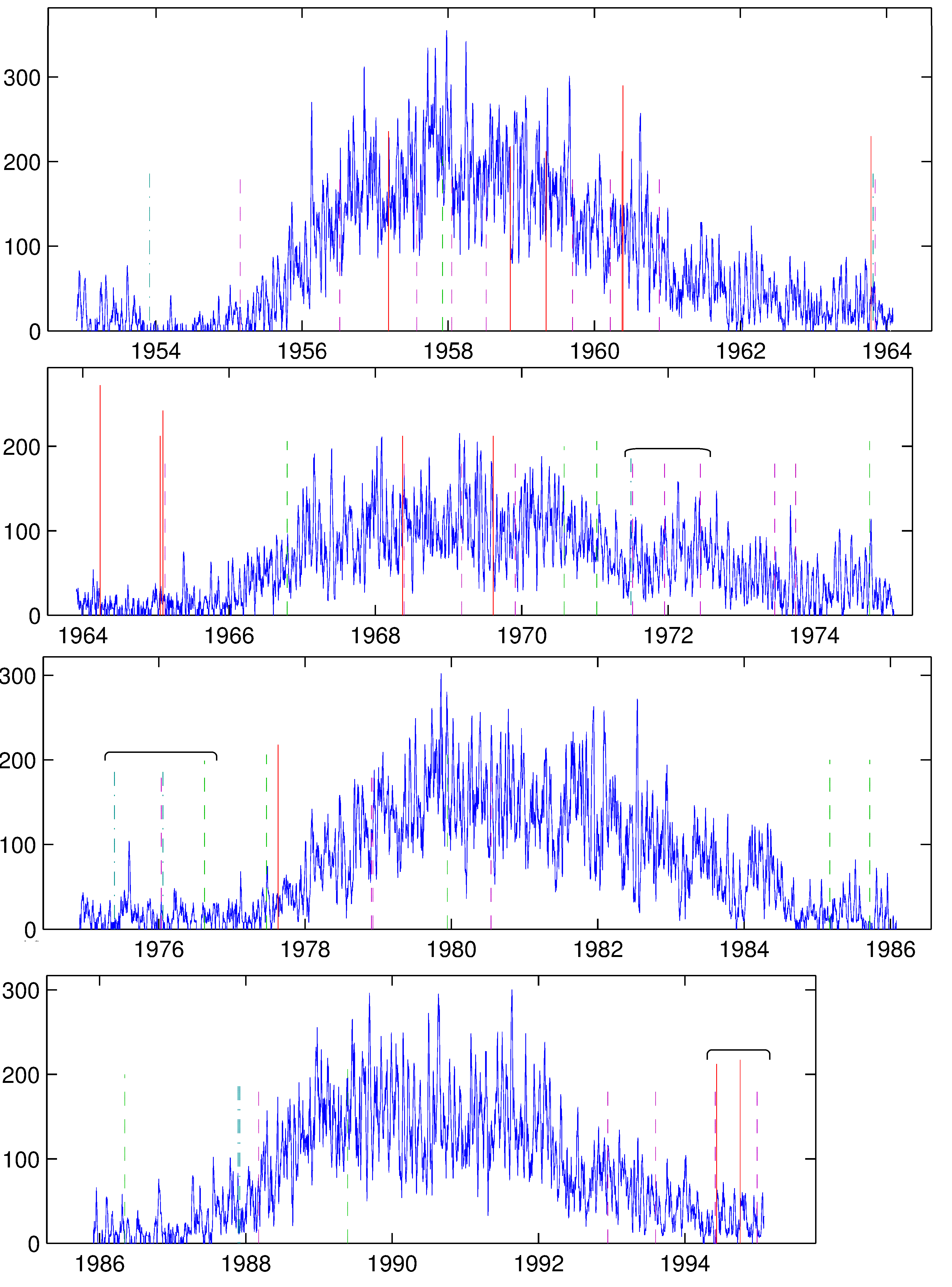} }
  \caption{ SSN vs earthquakes August 16, 1942 to December 31, 1985. Notations are explained in the caption to Figure \ref{fig:SSNvsEarthq3}.
  \hskip170mm  }
\label{fig:SSNvsEarthq2}
\end{figure}

\begin{figure}[!t]
  \centerline{\includegraphics[scale=0.14]{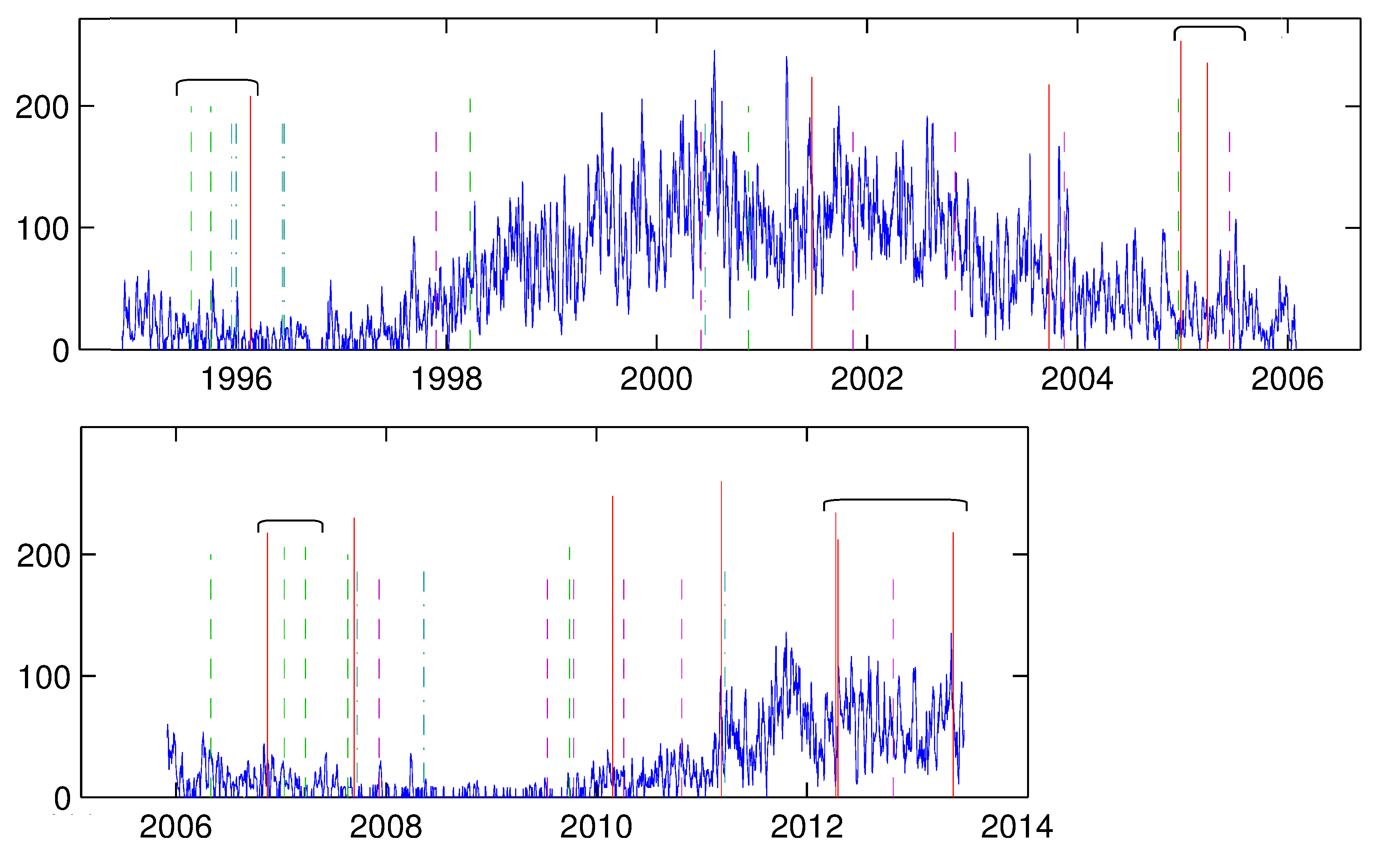} }
  \caption{ SSN vs earthquakes from January 1, 1986 to June 30, 2013. Red solid vertical lines represent  earthquakes of magnitude $> 8.1$,  green broken vertical lines represent  earthquakes of magnitude between $8.0$ and $8.1,$  green dot-dash vertical lines represent  earthquakes of magnitude   $7.9,$ and purple broken vertical lines represent  earthquakes of magnitude   $7.8;$  the length of each vertical line represents the magnitude of the corresponding earthquake. The blue curve is daily SSN. Horizontal brackets cover either groups of earthquakes with the time between the adjacent elements of the group to be about the same, or sets of earthquakes containing such groups.  The list of the earthquakes was obtained by taking all earthquakes of magnitude $\geqslant 8.0$  from \cite{largestearthq8} and supplementing with the  magnitude $7.8 -7.9$ from \cite{earthusgs}. Although both web sites are  produced by USGS,
some of the earthquakes are assigned different magnitude and/or different date, the differences are usually insignificant although in some cases they are fairly considerable, e. g.  \cite{largestearthq8} as of January 20, 2015 states that the July 31, 1970 earthquake in Columbia was of magnitude 8.0 and the January 10, 1971 earthquake in Indonesia was of magnitude 8.1 while \cite{earthusgs} states that the magnitudes of the two earthquakes were correspondingly 7.5 and 7.7.   Sifting through the two lists, double entries,  relatively insignificant  aftershocks and foreshocks were removed, dates and magnitudes were adjusted based on additional sources.
  \hskip170mm  }
\label{fig:SSNvsEarthq3}
\end{figure}

In Figures \ref{fig:SSNvsEarthq1}, \ref{fig:SSNvsEarthq2}, \ref{fig:SSNvsEarthq3} we superimpose magnitude $\geqslant 7.8 $ earthquakes on the graphs of SSN. We see the presence of earthquakes groups  just like in Figure \ref{fig:CRIvsEarthq}. The earthquakes are much less frequent when SSN increases  and CRI decays. There are two earthquakes between 1915 - early 1916, three earthquakes in late 1935 - early  1936, one earthquake in 1966, one earthquake in 1988, one earthquake in 1998 and a double earthquake in the beginning of  2011  which appear at the very beginning of   increasing stages of SSN; yet due to the lag of SSN behind CRI these earthquakes are at   maxima of CRI rather than  the decreasing stages. There are three earthquakes in late 1935 - early 1936,  one earthquake in 1956 and one in 1978 at   increasing stages of SSN yet SNN drops for 3-4 months prior to each earthquake exhibiting  behavior of being at a decreasing rather than increasing stage of a cycle.

\begin{figure}[!t]
  \centerline{\includegraphics[scale=0.25]{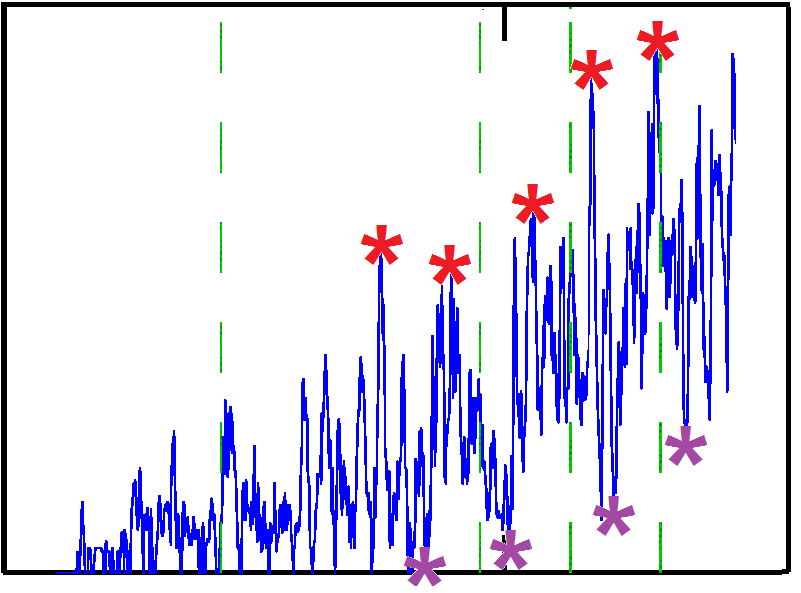} }
  \caption{ Part of the last frame in Figure \ref{fig:SSNvsEarthq1} showing  SSN vs earthquakes between the summer of 1945 and the summer of 1946. Although SSN was overall rising, it was also oscillating between the local maxima shown by red asterisks and the local minima shown by purple asterisks, the average frequency of oscillations was about the same as the average frequency of earthquakes of 129 days.
  \hskip170mm  }
\label{fig:SSNvsEarthq40s}
\end{figure}

The four earthquakes:  November 27, 1945, Pakistan; April 1, 1946, Alaska;  August 4, 1946, Hispaniola;  December 21, 1946,  Japan  are somewhat of a mystery. The first one was at a maximum of CRI and was preceded by a decrease in SSN/increase in CRI so it fits the pattern. However the other three were at an  increasing stage of SSN/decreasing stage of CRI whose presence cannot be easily explained: they are too far from the beginning of the SNN minimum/CRI maximum  to be explained   by  time lag nor were they  preceded by short-term decreases. These are the only three out of about 170 earthquakes, foreshocks and aftershocks of magnitude $\geqslant 7.8$ which appear at an increasing stage of SNN/decreasing stage of CRI.  What makes them even more mysterious is that the time intervals between the four earthquakes,   125, 125, and 139 days,  were almost the same, even though they were spread out all over the planet; the average time interval was about 130 days. As Figure \ref{fig:SSNvsEarthq40s} shows the solar activity, while on the rise overall, exhibited somewhat oscillatory behavior with the period of about  the same length. The years 1936-1951 saw an unusually large number of solar storms, in total 28, with almost two solar storms per year.  The earthquakes were preceded by a  sequence  of solar storms on 1) June 27, 1942;  2) September 4, 1943; 3) October 15  and   December 17, 1944 with the average date being November 16, 1944, it may be viewed as a single solar storm split up into two parts;  4) February 3, 1946. The number of days from June 27, 1942 to September 4, 1943 is 434;  from  September 4, 1943 to  November 16, 1944   is 439;  from  November 16, 1944 to February 3, 1946 is 444; the  average time  between two earthquakes was 439 days. The powerful solar storms undoubtedly created  sharp short-lived drops in CRI; could these almost periodic  drops in CRI, combined with Full Moon on June 28, 1942 and  New Moon on August 30, 1943,  October 17, 1944, November 15, 1944,   December 15, 1944 and and February 2, 1946,
have contributed  to the four earthquakes? Quite possible.

Below we consider several more examples of correlation between CRI and seismic activity on Earth.

\begin{figure}[!h]
 \centerline{\includegraphics[scale=.68]{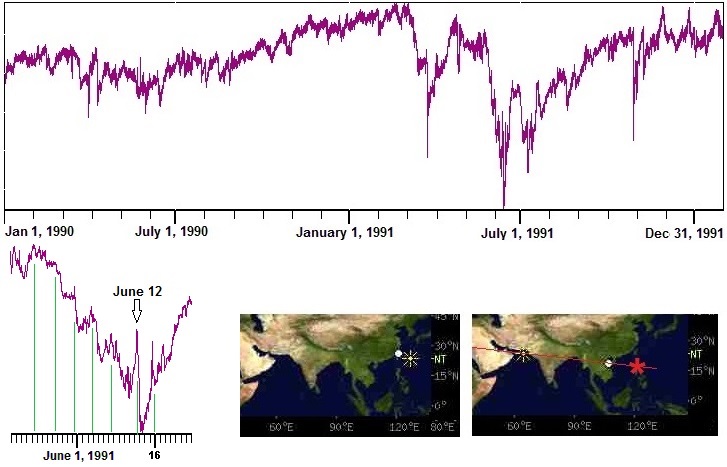}}
  \caption{ The top frame shows cosmic ray intensity  with resolution of one hour according to \protect \url{http://cr0.izmiran.rssi.ru/mosc/main.htm}, it shows a  sharp drops in     May-June,  1991.  The bottom left frame zooms in on the days of the   drop, the decreasing stage of the graph almsot periodic local maxima re-appearing every 4.8 days.  The  bottom center frame    shows that the sublunar and subsolar points were just above Mount Pinatubo on June 12, 1991 at 3:30 am UTC  just before the beginning of the violent phase of the eruption, according to   \protect \url{http://www.timeanddate.com/worldclock/sunearth.html?n=0&day=12&month=6&year=1991&hour=3&min=30&sec=0}. The bottom right frame   shows the sublunar and subsolar points on June 15, 1991 at 7:40 am UTC shortly before the climatic eruption, they were aligned with Mount Pinatubo shown by a red star, according to  \protect \url{http://www.timeanddate.com/worldclock/sunearth.html?n=0&day=15&month=6&year=1991&hour=7&min=40&sec=0};}
\label{fig:pinatubo}
\vskip2mm
 \centerline{\includegraphics[scale=.35]{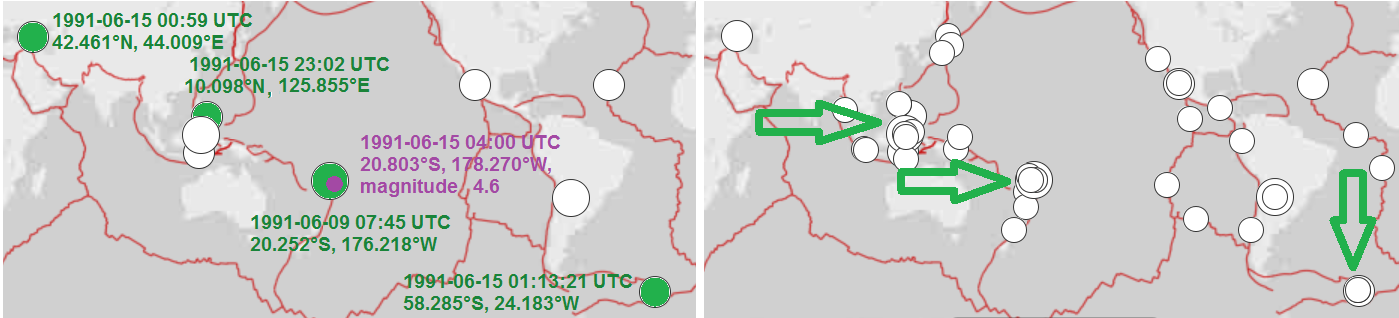}}
  \caption{The left frame shows earthquakes of magnitude $\geqslant6.0, $  the right frame shows earthquakes of magnitude $\geqslant 5.4.$  Three earthquakes marked  green in the left frame  are aligned along a straight line, the fourth one is very close to the line; on the Mercator projection all four are aligned along a straight line. Three of these earthquakes occurred on June 15, one on June 9 but there was a magnitude 4.6 earthquake at about the same location on June 15. Four earthquakes on the same day aligned along a straight line is quite unusual. Three of these earthquakes were accompanied by powerful foreshocks/aftershocks shown by green arrow in the frame on the right.  Source:  \protect \url{http://earthquake.usgs.gov/earthquakes/search/}.}
\label{fig:EarthqLineJune1991}
\end{figure}

 \noindent {\bf Example 1: three short-lived drastic drops in CRI. }  Figure \ref{fig:CRIvsEarthq} shows the drastic short-lived drop in CRI in May 25 - June 23, 1991, its bottom on June 13, 1991  is marked by I.  It coincided with a cataclysmic eruption of  	 Mount Pinatubo at $15.1417^{\rm o}N, 120.35^{\rm o}E$ in mid-June, 1991  after 500 years of dormancy, the second most powerful eruption of the 20th century; the violent phase of the eruption started on June 12 and achieved its climax on June 15.  The same day of  June 15 saw another remarkable event of four  earthquakes   along a single line,  as shown in Figure \ref{fig:EarthqLineJune1991}.  Just two days earlier on June 13,  CRI hit its lowest point ever, the Moon was New and at perigee. There was    a major solar storm on in the first week of June, 1991 and, according to \cite{solarflarejune1991}, a major solar flare on June 15, which, undoubtedly, contributed to the drop in CRI; yet   the drop in CRI began on May 24, more than a week prior to the solar activity and thus could not have resulted from solar activity alone.  Most likely, the drop in CRI contributed to the increase in solar activity which, in turn, exacerbated the drop in CRI.
The short-lived jump in CRI on June 12 shown in Figure \ref{fig:pinatubo}  coincided with typhoon Yunya which developed from a mild tropical disturbance on the same day.  The events were
 preceded by  smaller drop in CRI in   March - April and
 a succession of earthquakes  in March 15 - April 2, 1991.
  Yet,  as Figure \ref{fig:CRIvsEarthq} shows, there were no  powerful earthquakes  for a considerable time period before and after the drop in CRI.   A powerful magnitude 7.8 earthquake  on July 16, 1990 close to the site of Mount Pinatubo indicates that there was already seismic activity in the region prior to the eruption, the   drop in CRI, combined with the increased gravitational pull due to New Moon, might have been the "last straw" leading to the eruption.

 The second most drastic short-lived drop in CRI  occurred in July 11-20, 1982, its bottom is marked by G.   preceded by a somewhat slower drop in CRI in May-June, 1982.  It was only 1-2 months away from  three powerful   eruptions in March 29 - April 4   of volcano El Chich\'{o}n  in Mexico at $17.36^{\rm o}, 93.23^{\rm o}W$ after about 600 years of dormancy, although there are some debates about an eruption   around 1850.   As Figure \ref{fig:CRIvsEarthq} shows, again the drop in CRI was accompanied by  a long-term hiatus of  powerful earthquakes.

 The third largest drop in CRI was between October 27, 2003 and November 5, 2003, its bottom is between K and L. It was amidst of a periodic group of earthquakes of magnitude $\geqslant 7.8$\footnote{USGS seems to have recently revised its data with some of these earthquakes' magnitude changed a bit. }:  1) three earthquakes in November 16-17, 2000 in Papua New Guinea; 2) a double earthquake  on November 14, 2001 in China; 3) an earthquake  on November 3, 2002 in Alaska; 4) an earthquake on November 17, 2003 in Alaska; 5) four   magnitude $\geqslant 7.8$ earthquakes in December 23-28, 2004 in Indonesia.  The remarkable annual periodicity of the group suggests that all of its earthquakes had the same cause(s) initiated in 2000.  The drop  in CRI was accompanied by a single almost periodic group of   magnitude $\geqslant 7.8$ earthquakes with pre-2001 cause(s) and only June 23, 2001 and September 25,  2003 earthquakes besides the group, again showing a hiatus, magnitude $\geqslant 7.8$ earthquakes took at the time.

  \noindent {\bf Example 2:  short-lived drops in CRI in 1959 and 2000.}  A significant short-lived drop in CRI in the end of 1959 was followed by the largest known earthquake on May 22, 1960.  Another significant short-lived drop  in CRI occurred in July 2000, it was followed by a powerful eruption of volcano Ulawun on September 29, 2000 and a series of powerful earthquakes.

\begin{figure}[!h]
  \centerline{\includegraphics[scale=.25]{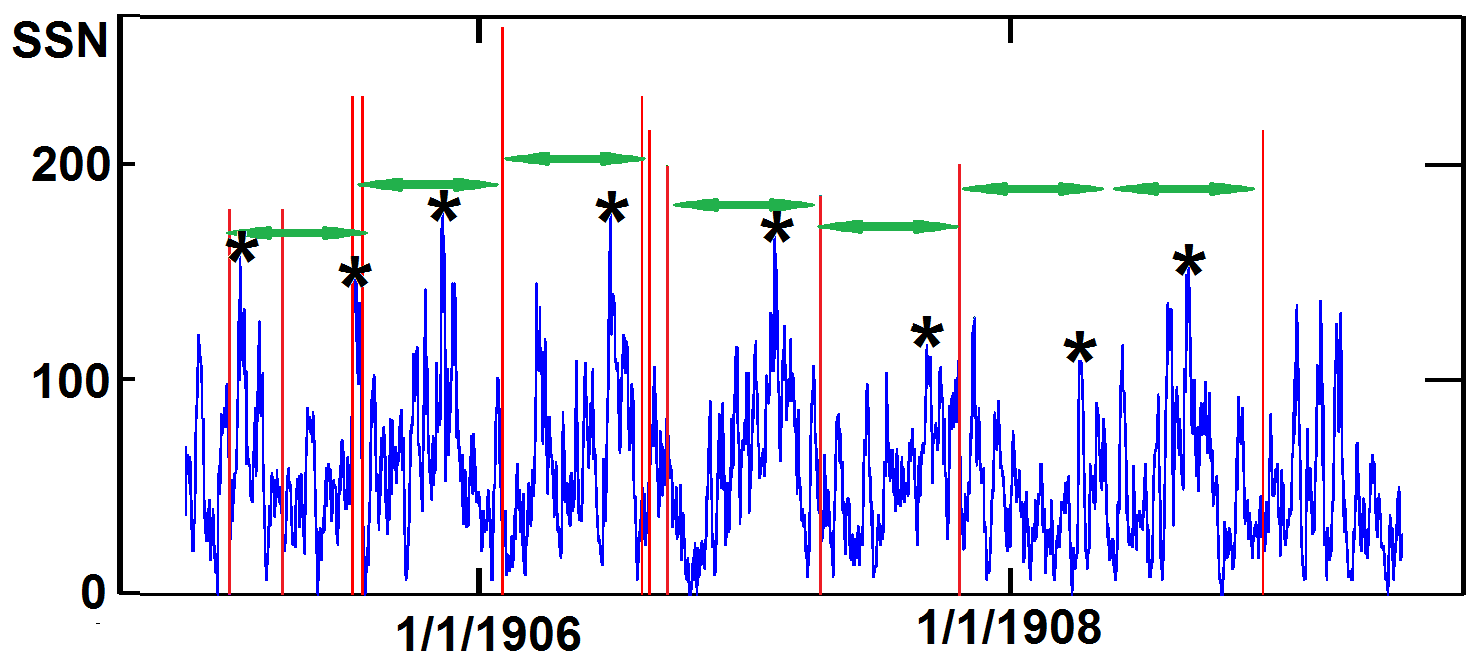}}
  \caption{   Sunspots vs earthquakes of magnitude $\geqslant 7.8$ in 1905-1910.  The graph shows SSN in blue color and  7 groups of earthquakes of magnitude $\geqslant 7.8$  marked by red vertical lines: 1) 1905/1/22, Indonesia, magnitude 7.8;  1905/4/4, India; magnitude 7.8;
  2)  1905/7/9, Mongolia,   magnitude 8.4;    1905/7/23, 	Mongolia,   magnitude 8.4; 	 3) 1906/1/31, Ecuador-Columbia, M=8.8;
  4)  1906/8/17, 	Chile,   magnitude  8.2;  906/8/17, Alaska, magnitude 8.4;    1906/9/15, Papua New Guinea, magnitude 8.0;
5) 1907/4/5,   Mexico, magnitude  7.9;  6)
  1907/10/21, Afghanistan,  magnitude 8.0;
7)  1908/12/12,   Peru,  	 magnitude  8.2. The   time interval between  groups 1-6 is the   number of days between 1905/1/22 and 1907/10/21 divided by 5 which is about 200 days, while  the number of days  between group  6, or 1907/10/21,  and group 7, or 1908/12/12,  is    419. The black asterisks mark maxima of SSN corresponding to short-term oscillations about every 200 days, all but the second last are followed by a group of earthquakes.
 }
\label{fig:SSNvsEarthq1900to1910}
\end{figure}

  \noindent {\bf Example 3: years 1906-1908. }  As Figure \ref{fig:SSNvsEarthq1} shows 1906-1908  were the years of a solar maximum. yet   SSN oscillated  up and down, as shown in Figure \ref{fig:SSNvsEarthq1900to1910}.   Six groups of powerful earthquakes struck almost periodically about every 200 days. Each group was preceded by a short-lived maximum of SSN oscillations with about the same period of 200 days.   Since, as pointed out earlier and explained in Figure \ref{fig:tm12}, CRI  minima lag  SSN maxima, the minima of CRI corresponding to these oscillations were at about the same time as the groups of powerful earthquakes.     Curiously, there was a seventh group  comprised of a single earthquake and separated from the sixth group by 419 days; it was  preceded by two maxima of short-term oscillations suggesting that there should have been an earthquake some time in June-July but there was none. Instead, there was a mysterious explosion near the river of  Tunguska in Siberia.

\begin{figure}[!h]
  \centerline{\includegraphics[scale=0.7]{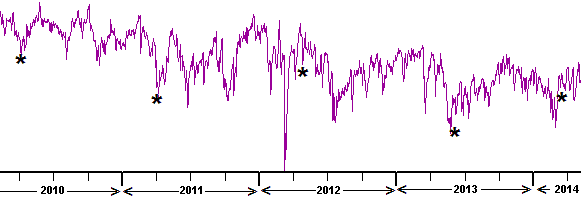} }
  \caption{ Daily average for CRI for 2010-2014. The asterisks approximately indicate the  points on the graph at the time of the earthquakes in Example 4. }
\label{fig:2010to2014}
\end{figure}

 \noindent {\bf Example 4:  earthquakes of magnitude  $\pmb \geqslant \pmb 8.2$  in 2010-2014. }  In 2010-2014 the earthquakes of magnitude $\geqslant 8.2$ struck on the following dates:   1) two earthquakes of magnitude 8.8 on February 27, 2010 at 6:34 UTC in Chile at $36.122^{\rm o}S, 72.898^{\rm o}W$  and     $35.85^{\rm o}S,  72.71^{\rm o}W;$  2) magnitude  9.0  earthquake on March 11, 2011 in   Japan at $38.297^{\rm o}N, 142.373^{\rm o}E,$ with the epicenter almost antipodal  to the earthquakes in the previous group; 3)  magnitude 8.6 earthquake on April 11, 2012 in   Indonesia at $    2.327^{\rm o}N , 93.063^{\rm o}E,$  followed by a magnitude 8.2 aftershock, the epicenters of both were almost antipodal to the earthquakes in the first group;  and 4) magnitude 8.3  earthquake on May 24, 2013 in the    Okhotsk Sea at $54.892^{\rm o}N, 153.221^{\rm o}E;$ 5) magnitude 8.2  earthquake on April 1, 2014 in Chile at $19.610^{\rm o}S, 70.769^{\rm o}W$ close to the epicenter in the first group and almost antipodal to the earthquake in the third group.  The earthquake in the Okhotsk Sea also had its antipodal counterpart; although  it was comprised of two less powerful   earthquakes, one of magnitude 7. 3 on July 15,  2013 at
    $60.857^{\rm o}S, 25.07^{\rm o}W,$   the other one of magnitude 7.7 on  November 17,   2013 at $60.274^{\rm o}S, 46.401^{\rm o}W.$
  The average  time interval  between   adjacent  earthquakes  was 373 days.
   Each one of these earthquakes occurred close to a CRI  minimum    as shown in Figure \ref{fig:2010to2014}.   The portion of Figure \ref{fig:CRIvsEarthq} showing  a much more smoothed graph of CRI for the period of 2010-2014,  shows almost the same periodicity  as the last five red lines denoting earthquakes of magnitude 8.2 or higher; each such earthquake  appears at the time or right after a sharp decrease in the graph of CRI.  The corresponding portion of the graph of SSN shows the same periodicity but it is far less pronounced than in the graph of CRI further suggesting that it is CRI rather than SSN that is correlated with earthquakes.

   \section*{  \centering{  Conclusion.}}

The discussion here here seems to point to conclusion that cosmic rays play much more prominent role that is currently believed; specifically: 1) cosmic ray  intensity seems to correlate with seismic activity on Earth much better than solar activity; 2) not only the solar activity regulates the flow of cosmic rays, as is currently accepted, but also the cosmic rays influence the solar activity, which currently is somewhat of a heretic statement.

  \section*{   Acknowledgments.  }

 The pictures/graphs used in this article were either reproduced from public domains with references provided or with permission of the creator(s), the authors would like to express their gratitude to all those who created the pictures and those who made them available.

 The data for the cosmic ray intensity are from \cite{cri}, the data for the solar spot numbers are from \cite{ssnNOAA, ssnSILSO}, the earthquake data   are from   \cite{largestearthq8}, \cite{earthusgs}, and  the Moon phases  are from \cite{fullmoon2} and \cite{apoperigee}.
Figures \ref{fig:CRIvsSSN},   \ref{fig:hath},    \ref{fig:hath2} are reproduced correspondingly  from \cite{climate4u},   \cite{hath},   \cite{hath2}; Figure    \ref{fig:carb} is reproduced   from \cite{delaygue} and  \cite{c14};
the reliability of the data is trusted to the creators of the graphs.

Although we are not aware of any previous work discussing any  correlation between earthquakes and CRI, there are numerous references on the correlation  between   earthquakes and SSN, for example   \cite{yum, anag}, many more may be found by typing appropriate expressions in an Internet search engine.

\bibliographystyle{plain}
\bibliography{SolarActivity}{ }
\end{document}